\documentclass[conference,a4paper]{IEEEtran}
\makeatletter
\def\ps@headings{%
\def\@oddhead{\mbox{}\scriptsize\rightmark \hfil \thepage}%
\def\@evenhead{\scriptsize\thepage \hfil \leftmark\mbox{}}%
\def\@oddfoot{}%
\def\@evenfoot{}}
\makeatother
\usepackage{simplemargins}
\settopmargin{0.92in}
\setbottommargin{0.5in}
\setleftmargin{0.568in} \setrightmargin{0.768in}
\usepackage{graphicx}
\usepackage{mathrsfs}
\usepackage{indentfirst}
\usepackage{epsfig}
\usepackage{amsfonts}
\usepackage{amssymb}
\usepackage{array}
\usepackage{color}
\usepackage{stmaryrd}
\usepackage{algorithm}
\usepackage{algorithmic}
\usepackage{subfigure}
\usepackage{amsmath}
\usepackage{caption}
\usepackage{epstopdf}
\usepackage{epsfig}
\usepackage{url}
\usepackage{bm}
\usepackage{fancyhdr}

\IEEEoverridecommandlockouts

\title{Multicast-aware Caching for Small Cell Networks}

\author{\IEEEauthorblockN{Konstantinos Poularakis, George Iosifidis, Vasilis Sourlas and Leandros Tassiulas}
\vspace{-4mm}
\IEEEcompsocitemizethanks{
\IEEEcompsocthanksitem
The authors are with the Department of Electrical and Computer Engineering, University of Thessaly, Volos, Greece. Emails: \{kopoular, giosifid, vsourlas, leandros\} @uth.gr
}
}

\begin{document}

\maketitle
\pagenumbering{arabic}

\newtheorem{property}{Property}
\newtheorem{proposition}{Proposition}
\newcommand{\be}{\begin{itemize}} \newcommand{\ee}{\end{itemize}}
\newcommand{\tb}{\textbf} \newcommand{\ttt}{\texttt}
\newcommand{\tit}{\textit} \newcommand{\uline}{\underline}
\newcommand{\argmin}{\operatornamewithlimits{argmin}}
\newcommand{\argmax}{\operatornamewithlimits{argmax}}
\newtheorem{theorem}{Theorem} \newtheorem{lemma}{\bf Lemma}
\setcounter{page}{1}
\thispagestyle{empty}

\begin{abstract}
The deployment of small cells is expected to gain huge momentum in the near future, as a solution for managing the skyrocketing mobile data demand growth. Local caching of popular files at the small cell base stations has been recently proposed, aiming at reducing the traffic incurred when transferring the requested content from the core network to the users. In this paper, we propose and analyze a novel caching approach that can achieve significantly lower traffic compared to the traditional caching schemes. Our cache design policy carefully takes into account the fact that an operator can serve the requests for the same file that happen at nearby times via a single multicast transmission. The latter incurs less traffic as the requested file is transmitted to the users only once, rather than with many unicast transmissions. Systematic experiments demonstrate the effectiveness of our approach, as compared to the existing caching schemes.
\end{abstract}


\section{Introduction} \label{section:1}

Today we are witnessing an unprecedented worldwide growth of mobile data traffic that is expected to surpass 10 exabytes per month in 2017 \cite{cisco}. In order to manage this load, operators deploy small cell base stations (SCBSs) that work in conjunction with the conventional macrocellular base stations (MBS). The SCBSs increase the area spectral efficiency and serve the users with short range energy-prudent transmission links. The main drawback of this approach however is the high cost incurred by the deployment of the backhaul links that connect the SCBSs to the core network \cite{intel}.

Local caching of popular files at the SCBSs has been recently proposed \cite{femto-dimakis}-\cite{facility-poularakis}, so as to reduce the necessary capacity, and hence the cost, of these backhaul links. Based on this novel architecture, user requests are served by the SCBSs, if the latter have cached the respective file, otherwise the MBS is triggered to serve them. The main challenge here is to design the optimal caching policy, i.e., to determine the files that should be cached in each SCBS so as to minimize the cost for serving the requests. However, one aspect of the cellular networks that has not been considered in these previous works, is that operators can employ multicast transmissions to concurrently serve multiple requests of different users.

Multicast constitutes a promising solution for efficient delivery of multimedia content over cellular networks (e.g., see \cite{korakis-multicast} and references therein), and has been incorporated in 3GPP specifications\footnote{The so-called multimedia broadcast multicast service (MBMS) was incorporated in Rel. 6, and more recently the respective enhanced version eMBMS in Rel. 9. Our analysis is generic and holds for other multicast/broadcast technologies, such as DVB-H, DVB-D, as well.} \cite{3gpp-multicast}. It can be used to deliver content to users that have subscribed to a multicast service, or to users that submit file requests at nearby times and hence can be served via a single multicast transmission. Clearly, multicast impacts the caching policies. For example, when MBS multicast is used to deliver a file, there is no need to cache it in any SCBS. On the other hand, in order to avoid such MBS transmissions, all the SCBSs that receive user requests for this file should have it cached.

In this paper, we design caching policies for small cell networks when the operators employ multicast. Similarly to previous works, our goal is to reduce the servicing cost of the operator by minimizing the volume of the incurred traffic. However, our approach differs substantially from other studies that designed caching policies based solely on content popularity \cite{femto-dimakis}, \cite{facility-poularakis}. Namely, multicast transmissions couple caching decisions with the spatiotemporal characteristics of user requests, and renders the problem NP-hard even for the simple case of non-overlapping SCBS coverage areas.

First, we demonstrate through simple examples how multicast affects the optimality of caching policies. Accordingly, we introduce a general optimization problem (which we name MACP) for devising the optimal caching policy under different user requirements. We assume that different users ask for different files in different time instances. The location and the time arrival of these requests determines whether MBS or SCBSs multicast transmissions are possible which, in turn, affects the design of the caching policies. We prove the complexity of the caching problem and provide a heuristic algorithm that yields remarkable results compared to conventional caching schemes.

Our main technical contributions are as follows:
\begin{itemize}
\item \emph{Multicast Aware Caching Problem (MACP)}. We introduce the MACP problem that derives caching policies which take into account the possibility of multicast transmission from MBS and SCBS. This is very important as content delivery via multicast is part of 3GPP standards and gains increasing interest.
\item \emph{Complexity Analysis of MACP}. We prove the intractability of the MACP problem by reducing it to the set packing problem \cite{NP}. That is, we show that MACP is NP-hard even to approximate within a factor of $O(\sqrt{N})$, where $N$ is the number of SCBSs.
\item \emph{Heuristic Solution Algorithm}. We present a heuristic algorithm that provides significant performance gains compared to the existing caching schemes. The problem formulation and the algorithm are generic in the sense that apply for general network parameters such as different servicing cost, coverage areas and user demands.
\item \emph{Performance Evaluation}. We evaluate the proposed scheme in representative scenarios. We show that our algorithm reduces the servicing cost even down to $52\%$ compared to conventional (multicast-agnostic) caching schemes and study the impact of several system parameters such as the cache sizes and the user request patterns.
\end{itemize}

The rest of the paper is organized as follows: Section \ref{section:2} reviews our contribution compared to the related works, whereas Section \ref{section:3} describes the system model and defines the problem formally. In Section \ref{section:4}, we show the intractability of the problem and present a heuristic caching algorithm with concerns on multicast transmissions. Section \ref{section:5} presents our numerical results, while in Section \ref{section:6} we conclude the paper.

\section{Related work} \label{section:2}

The idea of leveraging in-network storage for improving network performance is gaining increasing interest \cite{storage-ietf} and has been recently proposed also for small cell networks \cite{femto-dimakis}-\cite{facility-poularakis}. Authors in \cite{femto-dimakis} performed the file placement in storage capable base stations based solely on file popularity. The subsequent work in \cite{Layer-femto} extended their results for the special case that users request video files encoded into multiple quality levels. In our previous work \cite{facility-poularakis}, we studied the impact of SCBSs' wireless capacity constraints on the caching decisions. In contrast to all these studies, our caching policy is carefully designed with concerns on the multicast which is often used by operators to reduce the servicing cost. It is worth emphasizing that this twist increases significantly the complexity of the caching policy design problem. Namely, while for the simple scenario of non-overlapping coverage areas of the SCBS the conventional file placement is a trivial problem \cite{femto-dimakis}, we prove that incorporating multicast transmissions into the system makes it NP-hard.

The caching problem has also been studied in information delivery through broadcasting in conventional cellular networks (i.e., without SCBSs) \cite{Su-Tassiulas}. In these systems, users are endowed with caches in order to store in advance broadcasted content and retrieve later when they need it. The closest work to ours is that presented by Maddah-Ali et al. \cite{ISIT-multicast}. The authors focus on the joint caching and content delivery problem for the case that there exists a set of users, each one requesting a single file. The goal is to serve them with a single multicast transmission in a way that reduces the peak traffic rates. In contrast to that work, we consider a small cell network setting and aim at deriving the caching policy that minimizes the average cost incurred when serving the user requests.

\section{System model and problem formulation} \label{section:3}

In this section we introduce the system model, we provide a motivating example that explains the problem under consideration and highlights the impact of multicast on caching and, finally, we formally define the multicast-aware caching optimization problem.

\begin{figure}[t]
\begin{center}
\includegraphics[scale=0.48]{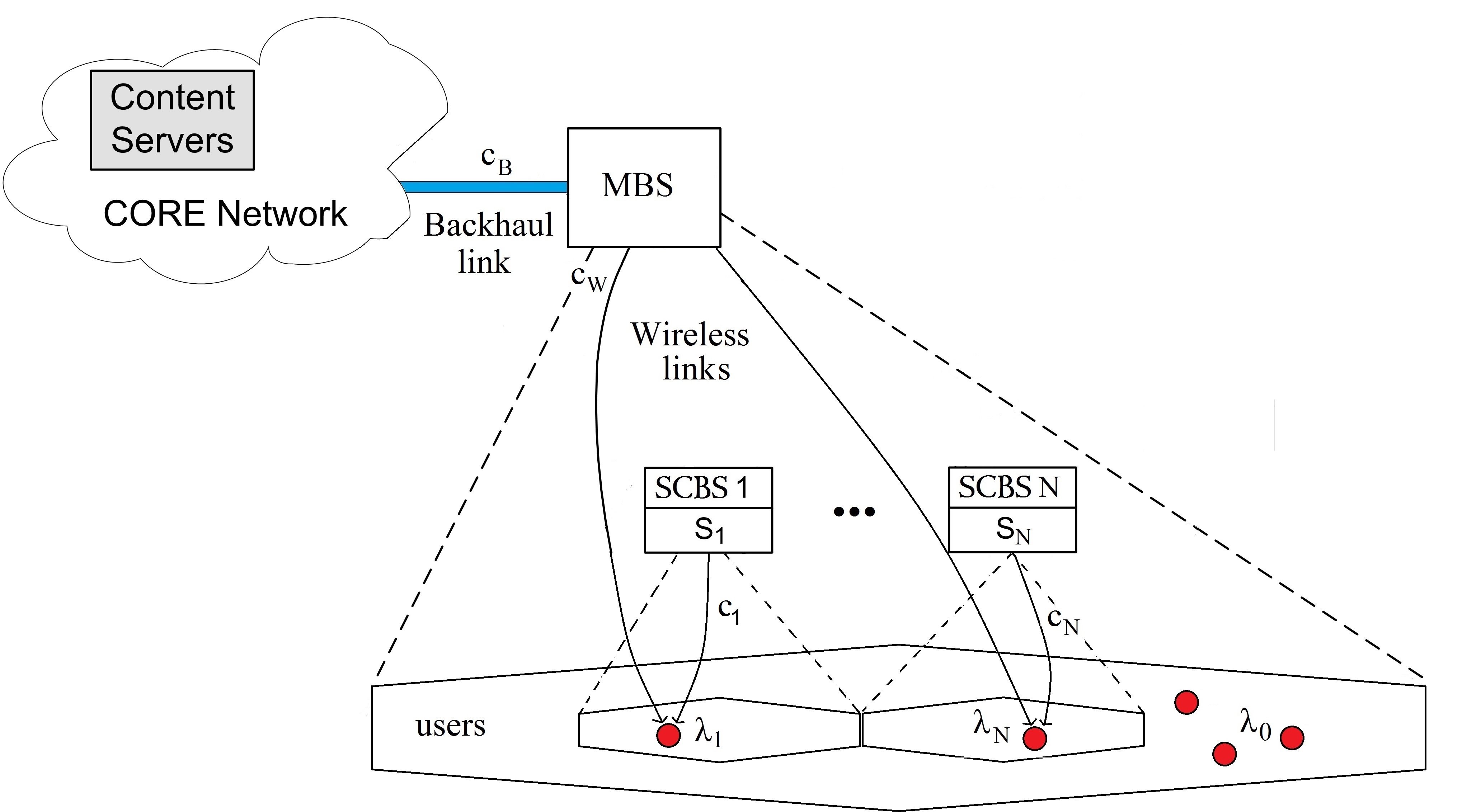}
\caption{Graphical illustration of the discussed model. The hexagons represent the coverage areas of the MBS and the SCBSs.
}
\label{fig:model}
\end{center}
\vspace{-4mm}
\end{figure}

\textbf{System Model}. We study the downlink operation of a small cell network like the one depicted in Figure \ref{fig:model}. A set $\mathcal{N}$ of $N=|\mathcal{N}|$ small cell base stations (SCBSs) are deployed within the macrocell\footnote{The model can be directly extended for the scenario of more macrocells where different cells employ different multicasts or coordinate via single-frequency network configurations \cite{3gpp-multicast}.}, serving the requests of the nearby users. Each SCBS $n\in\mathcal{N}$ is equipped with a cache of size $S_n\geq 0$ bytes.

The MBS is connected to the core network via a backhaul link. We denote with $c_B\geq 0$ the average incurred cost per byte (in monetary units/byte) when transferring data from the core network to the MBS via it's backahul link. Parameter $c_B$ refers to the average cost of the required backhaul capacity in case the link is owned by the operator. When the backhaul link is leased (e.g., by a Tier-1 ISP), $c_B$ denotes the average cost per byte paid to the link provider, which depends both on peak traffic and on the volume of the traffic, based on the employed pricing scheme. Besides, we denote with $c_W\geq 0$ the cost per byte incurred when transmitting data directly from MBS to the users in the cell. Parameter $c_W$ refers to the average MBS energy consumption when transmitting files to the users. Finally, let $c_n\geq 0$ denote the unit cost incurred when transmitting data from the SCBS $n$ to it's nearby users. Clearly, $c_n \leq c_W$, $\forall n \in \mathcal{N}$, since the SCBSs are in closer proximity to the users than the MBS. In general, the above cost parameters can be interpreted as the average OpEx, and average projected CapEx costs of the operator.

We study the system for a certain time interval (several hours or few days), during which the users demand for a set of popular files is assumed to be known in advance, as in \cite{femto-dimakis}, \cite{markov-poularakis}, \cite{facility-poularakis}. Let $\mathcal{I}$ indicate that collection of $I=|\mathcal{I}|$ content files. For notational convenience, we assume that all files have the same size normalized to $1$. This assumption can be easily removed as, in real systems, files can be divided into blocks of the same length \cite{femto-dimakis}, \cite{markov-poularakis}, \cite{facility-poularakis}. Users are heterogeneous since they may have different content demands.

To facilitate the analysis, we consider the case that the coverage areas of the SCBSs are non-overlapping, hence each user is in the coverage area of at most one SCBS. We want to emphasize at this point that, as it will be explained in the sequel, all the presented results for the complexity of the problem as well as the proposed algorithm hold also for the case of overlapping SCBS coverage areas. We denote with $\lambda_{ni} \geq 0$ the average demand of users for file $i$ covered by SCBS $n$ within the considered time interval. Also, $\lambda_{0i}\geq 0$ denotes the average demand of the users for file $i$ that are not in the coverage area of any of the SCBSs\footnote{Notice that, the current practice of operators is to deploy SCBSs to certain areas with high traffic. Hence, other less congested areas may be covered only by the MBS.}.

We assume that file requests must be satisfied within a given time deadline of $d$ seconds in order to be acceptable by the users, as in \cite{markov-poularakis}. The multicast service happens every $d$ seconds, which ensures that all the requests will be satisfied within the time deadline. We denote with $p_{nid}$ the probability that at least one request for file $i$ is generated by users in the coverage area\footnote{With a slight abuse of notation we use the same index for base stations and their coverage area.} of SCBS $n$ (area $n$) within the time period $d$. Similarly, $p_{0id}$ denotes the respective probability for the users that are not in the coverage area of any of the SCBSs (area $n_0$). We denote with $p(r,i,d)$ the probability that at least one request for the file $i \in \mathcal{I}$ is generated within each one of the areas $r \subseteq \mathcal{N} \cup\{n_0\}$, within the period $d$.

The operator can employ multicast to simultaneously serve many different requests for the \emph{same file} that happen within the \emph{same time interval} of duration $d$, and thus reduce its servicing cost. We assume that both SCBSs and MBS can use multicast. Namely, each SCBS $n\in\mathcal{N}$ multicast transmissions satisfy the requests of users within its coverage area, while MBS transmissions satisfy requests generated within the coverage areas of different SCBSs (and requests from area $n_0$) that have not cached the requested file. This latter option induces higher cost since the MBS has higher transmission cost and also needs to fetch the file via its backhaul link. This exactly is the main idea of this work: \emph{``To carefully design the caching policy with concerns on the multicast transmissions so as to minimize the servicing cost''}.

Before we introduce formally the problem, let us provide a simple example that highlights how the consideration of multicast transmissions impacts the caching policy.

\begin{figure}[t]
\begin{center}
\includegraphics[scale=0.09]{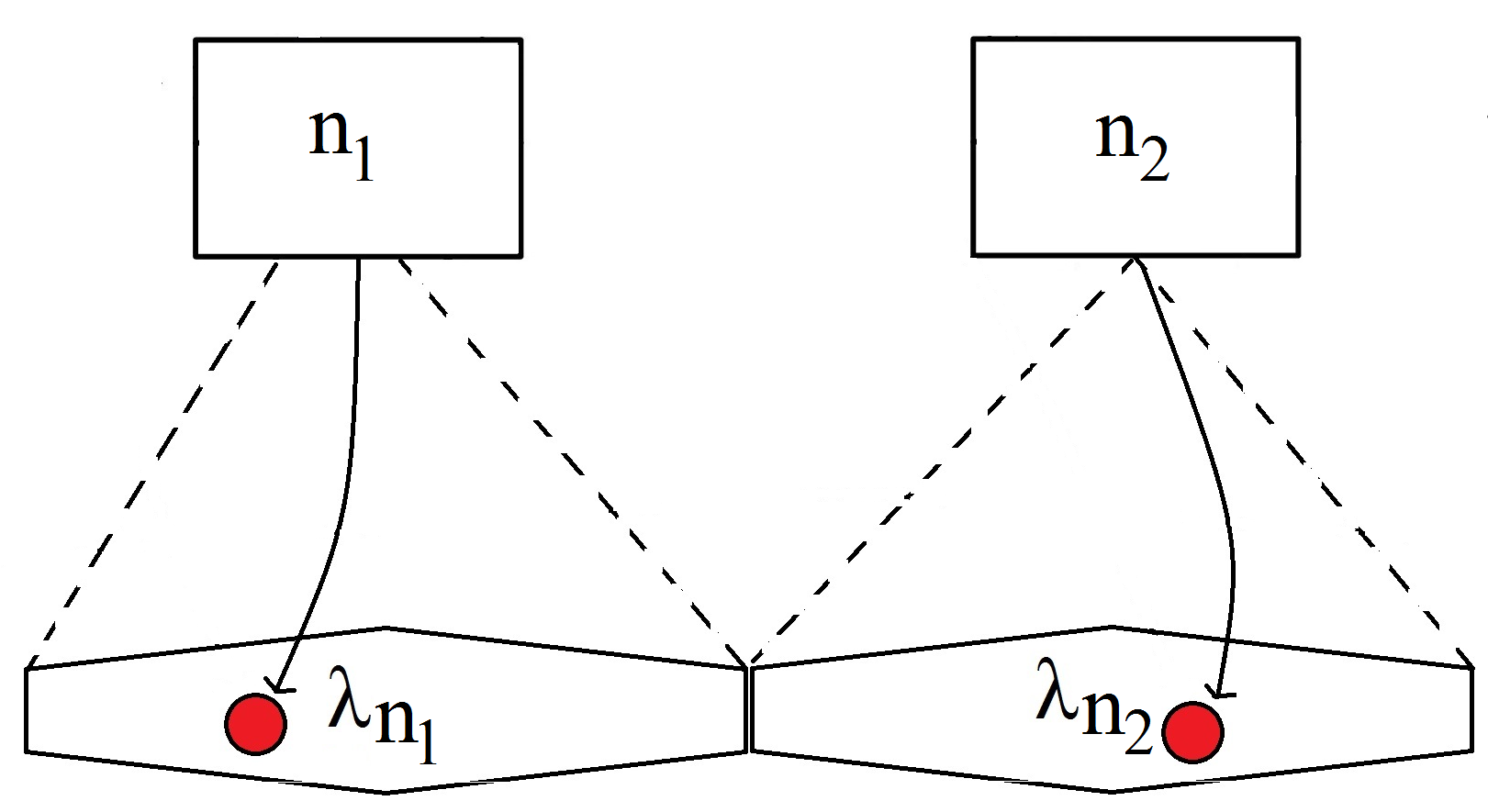}
\caption{An example with two SCBSs ($n_1$ and $n_2$).}
\label{fig:example}
\end{center}
\vspace{-6mm}
\end{figure}

\textbf{Motivating Example}. Consider the scenario depicted in Figure \ref{fig:example} with two SCBSs ($n_1$ and $n_2$). There are three equal sized files ($i_1$, $i_2$ and $i_3$). Each SCBS can cache at most one file because of it's limited cache size. We also set $c_B+c_W=1$, $c_{n_1}=c_{n_2}=0$ and $d=1$. We assume that requests are generated independently among different areas. Thus, for each subset of areas $r \subseteq \mathcal{N} \cup \{n_0\}$ it holds that:
\begin{equation}
p(r,i,d)=\prod_{n \in r}(p_{nid})\cdot \prod_{n \notin r}(1-p_{nid})\label{eq:product}
\end{equation}
Besides, we assume that the number of requests for each file $i$ within the coverage of SCBS $n$ follows a Poisson probability distribution with rate parameter $\lambda_{ni}$, $\forall n\in\mathcal{N},i\in\mathcal{I}$. Thus, the probability that at least one request for file $i$ is generated within SCBS $n$ in the time period $d$ is:
\begin{equation}
p_{nid} = 1 - e^{-\lambda_{ni} d}
\end{equation}
Let $\lambda_{n_1i_1}=0.51$, $\lambda_{n_1i_2}=0.49$, $\lambda_{n_1i_3}=0$, $\lambda_{n_2i_1}=0.51$, $\lambda_{n_2i_2}=0$, and $\lambda_{n_2i_3}=0.49$. Then, $p_{n_1i_11}=0.3995$, $p_{n_1i_21}=0.3874$, $p_{n_1i_31}=0$, $p_{n_2i_11}=0.3995$, $p_{n_2i_21}=0$ and $p_{n_2i_31}=0.3874$. The optimal caching policy places $i_2$ to $n_1$ and $i_3$ to $n_2$. Then, all the requests for $i_1$ will be served by transferring it via the backhaul link that connects the core network to the MBS and then transmitting it by the MBS (via a single multicast). The requests for the rest files will be satisfied by the accessed SCBSs (at zero cost). Hence, the total servicing cost is: $(c_B+c_W)\cdot\big(p_{n_1i_1}\cdot(1-p_{n_2i_1}) + (1-p_{n_1i_1})\cdot p_{n_2i_1} + p_{n_1i_1}\cdot p_{n_2i_1}\big)=0.6394$.

However, if we ignore the multicast transmissions for aggregated requests when designing the caching policy, (and thus assume that each request will be served via a separate unicast transmission), then \emph{the optimal caching policy changes}; it places file $i_1$ to both SCBSs (because $i_1$ is the most popular file according to $\lambda_1$ and $\lambda_2$). Then, the requests within $n_1$ for $i_2$ and the requests within $n_2$ for $i_3$ will be served by the MBS. The total servicing cost is: $(c_B+c_W)\cdot p_{n_1i_2}\cdot(1-p_{n_2i_3}) + (c_B+c_W)\cdot(1-p_{n_1i_2})\cdot p_{n_2i_3} + 2\cdot(c_B+c_W)\cdot p_{n_1i_2}\cdot p_{n_2i_3}= 0.7747>0.6394$, where the last term in the summation is multiplied by $2$ because \emph{two different files are requested for download} and thus can not be served with a single multicast transmission (i.e., \emph{two unicast transmissions are required}). This example demonstrates that ignoring the multicast transmissions in cache management decisions fails to fully exploit the multicast opportunities, and hence yields increased network servicing cost.

\textbf{Problem Statement}.
Let us introduce the integer decision variable $x_{ni} \in \{0, 1\}$, which indicates whether file $i \in \mathcal{I}$ is
placed at the cache of SCBS $n \in \mathcal{N}$ or not. We also define the respective caching policy matrix $\bm{x} = (x_{ni} : n \in \mathcal{N}, i \in \mathcal{I})$. To facilitate notation we introduce variable $y(\cdot)$ which indicates whether a multicast transmission by the MBS will happen, for a given caching policy $\bm{x}$, and a subset of areas $r$ requesting file $i$:
\begin{equation}
y(\bm{x},r,i)= \max  \big( \max_{n \in r \setminus \{n_0\}}(1-x_{ni}), \bm{1}_{\{n_0 \in r\}} \big) \in \{0,1\},
\end{equation}
where $\bm{1}_{\{.\}}$ is the indicator function, i.e., $\bm{1}_{\{c\}}$ is equal to one iff condition $c$ is true; otherwise it is equal to zero. For example, if a request is generated in a point that is not in the coverage area of any SCBSs, i.e. $n_0 \in r$, then a multicast transmission will happen as the requester can not find the file at a SCBS cache. Thus, $\bm{1}_{\{n_0 \in r\}}=1$ and the external max term is equal to 1. Similarly, $y(\bm{x},r,i)=1$ for the case that a request is generated within the coverage area of \emph{at least one} SCBS $n \in r \setminus \{n_0\}$, but the latter has not stored in it's cache the requested file.

The problem of determining the caching policy that minimizes the total servicing cost can be written as follows:
\begin{eqnarray}
\min _{\bm{x}}  & \sum\limits_{i \in \mathcal{I}} \sum\limits_{r \subseteq \mathcal{N} \cup \{n_0\}, ||r|| \geq 1}  p(r,i,d) \cdot \Big( y(\bm{x},r,i) \cdot (c_B+c_W) &   \nonumber\\
&+\big(1-y(\bm{x},r,i)\big) \cdot \sum_{n \in r} c_n  \Big) &\label{eq:objective}\\
s.t. &  \sum_{i \in \mathcal{I}} x_{ni} \leq S_n, \text{ } \forall n\in\mathcal{N} &\label{eq:storeconstraint}\\
&   x_{ni} \in \{0, 1\}, \text{ } \forall n\in\mathcal{N},\,i\in\mathcal{I}&\label{eq:constraint}
\end{eqnarray}
The above expression in the objective function indicates that for each subset of areas $r$ that generate at least one request for file $i$, within the same time period of duration $d$, a single multicast transmission by the MBS happens, if there is at least one requester that is not in range with a SCBS having cached the file $i$. In other case, i.e., when  $y(\bm{x},r,i)=0$, all the requests are satisfied by the accessed SCBSs. Constraints (\ref{eq:storeconstraint}) denote the cache capacity constraints of the SCBSs, whereas inequalities (\ref{eq:constraint}) indicate the discrete nature of the optimization variables. We call the above the \emph{Multicast-Aware Caching Problem} (MACP).

Observe that the description of the objective function in (\ref{eq:objective}) is exponentially long in the number of SCBSs $N$ due to the number of subsets $r \subseteq \mathcal{N} \cup \{n_0\}$. In practice though, its description is affordable as the number of SCBSs $N$ in a single cell is typically small (e.g., a few decades). Even so, as we prove in the next section, MACP is an NP-hard problem.

\section{MACP Complexity and a Heuristic Algorithm} \label{section:4}

In this section we prove the high complexity of the MACP problem and present a heuristic algorithm for it's solution. Namely, we show that the MACP problem is NP-hard by proving that the well known set packing problem (SPP) is polynomial-time reducible to MACP. In other words, we prove that SPP is a special case of MACP. Since SPP is NP-hard it directly follows that MACP is also NP-hard. Therefore, the following theorem holds:
\begin{theorem}
\emph{MACP is an NP-hard problem. Moreover, it is NP-hard even to approximate it within a factor of
$O( \sqrt{N})$}.
\end{theorem}
\vspace{1mm}

In order to prove Theorem $1$ we will consider the corresponding (and equivalent) decision problem, called Multicast Aware Caching Decision Problem (MACDP). Specifically:

\vspace{1mm}
\emph{MACDP}: Given a set $\mathcal{N}$ of SCBSs, a set $\mathcal{I}$ of unit-sized files, the vector $\bm{S}=(S_n:n\in\mathcal{N})$, the costs $c_B, c_W$ and $\bm{c}=(c_n:n\in\mathcal{N})$, the time deadline $d$, the request probability matrix $\bm{p}=\big(p(r,i,d): \forall r \subseteq \mathcal{N} \cup \{n_0\}, i \in \mathcal{I}\big)$, and a real number  $Q \geq 0$, \emph{we ask the following question}: does there exist a caching policy $\bm{x}$, such that the value of the objective function in (\ref{eq:objective}) is less or equal to $Q$ and constraints (\ref{eq:storeconstraint})-(\ref{eq:constraint}) are satisfied? We denote this problem instance with $MACDP(\mathcal{N},\mathcal{I}, \bm{S}, c_B, c_W, \bm{c}, d, \bm{p}, Q)$.

The set packing decision problem is defined as follows:

\vspace{1mm}
\emph{SPP}: Consider a finite set of elements $\mathcal{E}$ and a list $\mathcal{L}$ containing subsets of $\mathcal{E}$. We ask: do there exist $k$ subsets in $\mathcal{L}$ that are pairwise disjoint? Let us denote this problem instance by $SPP(\mathcal{E}, \mathcal{L}, k)$.
\vspace{1mm}

\begin{lemma} The set packing problem is polynomial-time reducible to the MACDP.
\end{lemma}
\begin{proof}
Consider the $SPP(\mathcal{E}, \mathcal{L}, k)$ decision problem and a specific instance of MACDP with $N=|\mathcal{E}|$ SCBCs, i.e., $\mathcal{N}=\{1,2,\ldots,|\mathcal{E}|\}$, a file set of $I=|\mathcal{L}|$ unit-sized files, i.e., $\mathcal{I}=\{1,2,\ldots, |\mathcal{L}|\}$, unit-sized caches: $\bm{S}=(S_n=1:n\in\mathcal{N})$, $c_B=0$, $c_W=1$, and $\bm{c}=(c_n=0:n\in\mathcal{N})$. Parameter $d$ is any positive number, and the question is if we can satisfy the users requests with cost $Q=1-\frac{k}{|\mathcal{L}|}$, where $k$ is the parameter from the SPP. The important point is that we define the elements of matrix $\bm{p}$ as follows:
\vspace{-1.5mm}
\begin{equation}
p(r,i,d)=
\begin{cases}
1/|\mathcal{L}| & \text{if $r=\mathcal{L}(i)$},\\
0 & \text{else}.
\end{cases}
\label{eq:probs}
\end{equation}
Observe that for given SPP instances, we can construct the respective specific MACDP in polynomial time.

Notice that with the previous definitions, $\mathcal{L}(i)$ is the $i^{th}$ component of the list $\mathcal{L}$ and containts a certain subset of elements of $\mathcal{E}$. For the MACDP, under the above mapping, this correspond to a subset of SCBSs asking with a non-zero probability file $i\in\mathcal{I}$. Moreover, with (\ref{eq:probs}) we assume that these probabilities are equal and have value $1/|\mathcal{L}|$.

If the MBS has to serve all the requests, then the MACDP problem has a value (cost) of $1$ (the worst case scenario). For each file $i$ that the operator manages to serve completely through local caching at the SCBSs, the operator reduces its cost by $(c_B+c_w)\cdot p(r,i,d)=1/|\mathcal{L}|$. This reduction is ensured only if the file is cashed in all the SCBSs $n\in r$ for which $p(r,i,d)=1/|\mathcal{L}|$. Therefore, in order to achieve the desirable value $Q=1-\frac{k}{|\mathcal{L}|}$, we need to serve locally the requests for $k$ files. That is, to find subsets of SCBSs $r$ where each file $i$ should be cached so as to avoid MBS multicasts.

Notice now that the caches are unit-sized. Hence, the caching decisions should be disjoint with respect to the SCBSs.
For example, in Figure \ref{fig:NP}, you can not store in SCBS 1 both files 1 and 2, because $S_1=1$. This ensures that you will not pick both the subsets $\{1\}$ and $\{1,2\}$ in the SPP problem.
In other words, the value of the objective function in (\ref{eq:objective}) can be less or equal to $1-\frac{k}{|\mathcal{L}|}$, if there exist $k$ subsets in $\mathcal{L}$ that are pairwise disjoint.

\begin{figure}[t]
\begin{center}
\includegraphics[scale=0.111]{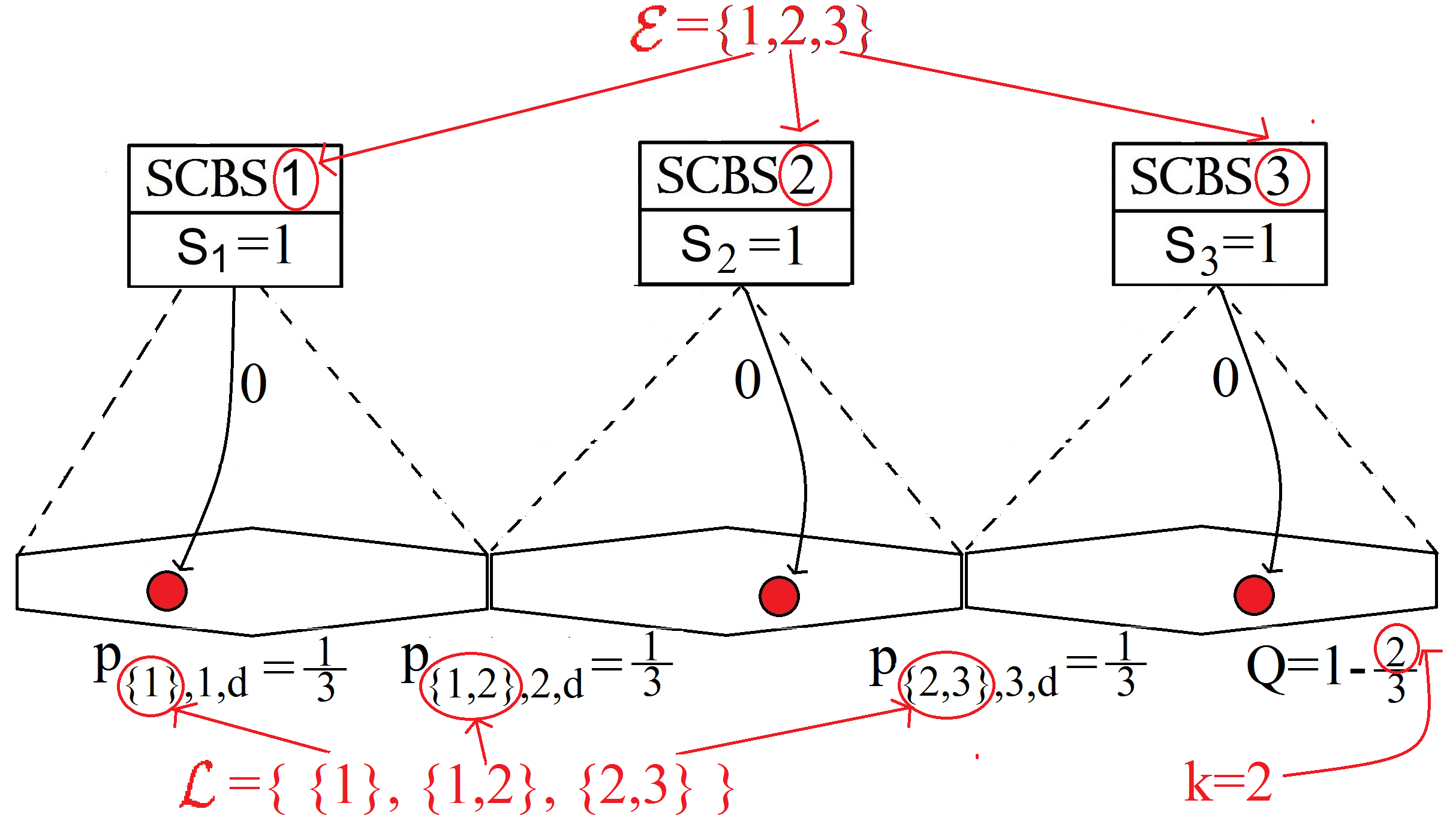}
\caption{An example of the reduction from $SPP(\{1,2,3\},\{\{1\},\{1,2\},\{2,3\}\},2)$.
In the MACDP instance there are N=3 SCBSs and I=3 files. There is a solution to MACDP
of cost $Q=1-\frac{2}{3}$ that places file $1$ to SCBS 1 and file $3$ to SCBSs 2 and 3.
Accordingly, the solution to SPP picks the subsets $\mathcal{L}(1)=\{1\}$ and $\mathcal{L}(3)=\{2,3\}$.}
\label{fig:NP}
\end{center}
\vspace{-7mm}
\end{figure}

Conversely, if a Set Packing for some $k$ exists, then for each subset $\mathcal{L}(i)$ that is picked in it, one can place the file $i$ to the cache of each one of the SCBSs $n\in \mathcal{L}(i)$ corresponding to this subset. At most one file is placed in each cache, since the picked subsets in the list are pairwise disjoint. The cost will be equal to $1-\frac{k}{|\mathcal{L}|}$.
\end{proof}

SPP is NP-hard and moreover it is inapproximable within  $O( \sqrt{|\mathcal{E}|})$ \cite{NP}. According to the reduction, we create a SCBS for each one of the elements in $\mathcal{E}$, and hence it holds $|\mathcal{E}|=N$, which concludes the proof of Theorem $1$.

At this point, we need to emphasize that Theorem $1$ holds also for the more general case that the SCBSs coverage areas are overlapping, which can be directly proved as this is a harder problem than the non-overlapping SCBSs that we considered in our analysis. This indicates that the multicast-aware problem is very hard even for the more simple non-overlapping coverage areas scenario.

\textbf{Heuristic Algorithm}. Because of the above hardness results, we propose a light-weight heuristic algorithm for the solution of the MACP problem. The proposed iterative algorithm starts with all the caches empty. At each iteration, it places the file to a non-full cache that yields the lowest value of the objective function in (\ref{eq:objective}). The algorithm terminates when all the caches become full. This is a greedy ascending procedure that can be summarized in Algorithm $1$.

\begin{algorithm}[t]
\caption{}
\begin{algorithmic}
\STATE INPUT: $\mathcal{N}$, $\mathcal{I}$, $d$, $S_n$, $\forall n \in \mathcal{N}$, $\lambda_{ni}$, $\forall n \in \mathcal{N}\cup \{n_0\}, i \in \mathcal{I}$
\STATE OUTPUT: The caching policy $\bm{x}$
\\\hrulefill
\STATE  $ \bm{x} \leftarrow  [0,...,0]$
\STATE $F_{n} \leftarrow 0$, $\forall n \in \mathcal{N}$
\STATE $\mathcal{D} \leftarrow \mathcal{N} \times \mathcal{I}$
\FOR{\emph{iteration}$=1,2,...,\sum_{n \in \mathcal{N}}(S_n)$}
\STATE $(n^{*}, i^{*}) \leftarrow argmin_{(n,i)\in \mathcal{D}} f(\bm{x},n,i) $
\STATE $x_{n^{*}i^{*}}=1$
\STATE $\mathcal{D}\leftarrow\mathcal{D} \setminus (n^{*}, i^{*})$
\STATE $F_{n^*} \leftarrow F_{n^*} + 1$
\IF {$F_{n^*}=S_{n^*}$}
\FOR{each $i \in \mathcal{I}$ such that $(n^*,i) \in \mathcal{D}$}
\STATE {$\mathcal{D} \leftarrow \mathcal{D} \setminus (n^*,i)$}
\ENDFOR
\ENDIF
\ENDFOR
\end{algorithmic}
\end{algorithm}

Specifically, $F_n$ is the number of files already placed at the cache of SCBS $n$ at every iteration of the algorithm, and $(\times)$ denotes the cartesian product of two sets. The set $\mathcal{D}$ includes all the pairs $(n,i)$ for which the placement of file $i$ at the cache of SCBS $n$ has not been performed yet, and the cache of $n$ has not been filled up yet.
$f(\bm{x},n,i)$ is the value of the objective function of the MACP for the file placement $\bm{x}$, where we additionally set $x_{ni}=1$ to evaluate the contribution of the $(n,i)$ pair. At every iteration, Algorithm 1 picks the pair $(n^{*},i^{*}) \in \mathcal{D}$ with the lowest cost value $f( \bm{x}, n^{*},i^{*} )$.
This corresponds to the placement of the file $i^{*}$ at the cache of the SCBS  $n^{*}$. If the cache of SCBS $n^{*}$ becomes full, then the algorithm excludes all the pairs $(n^{*},i)$, $\forall i$, from the set $\mathcal{D}$. That way, no more files will be placed at this cache. The algorithm terminates when all the caches become full.

Algorithm 1 requires $\sum_{n=1}^{N} (S_n) $ iterations until all the caches become full.
At each iteration it evaluates the value of the objective function after the placement of each one of at most $N \cdot I$ files.

To avoid confusion, we note here that Algorithm 1 is known to achieve an approximation ratio equal to 2 for the caching problem without multicast \cite{femto-dimakis}.
However, MACP problem is hard even to approximate below $O(\sqrt{N})$, as we proved in Theorem 1.
Even so, as we show numerically in the next section, Algorithm 1 provides significant performance gains compared to the existing caching schemes.

\begin{figure*}[t]
\centering
\subfigure{
    \includegraphics[scale=0.17]{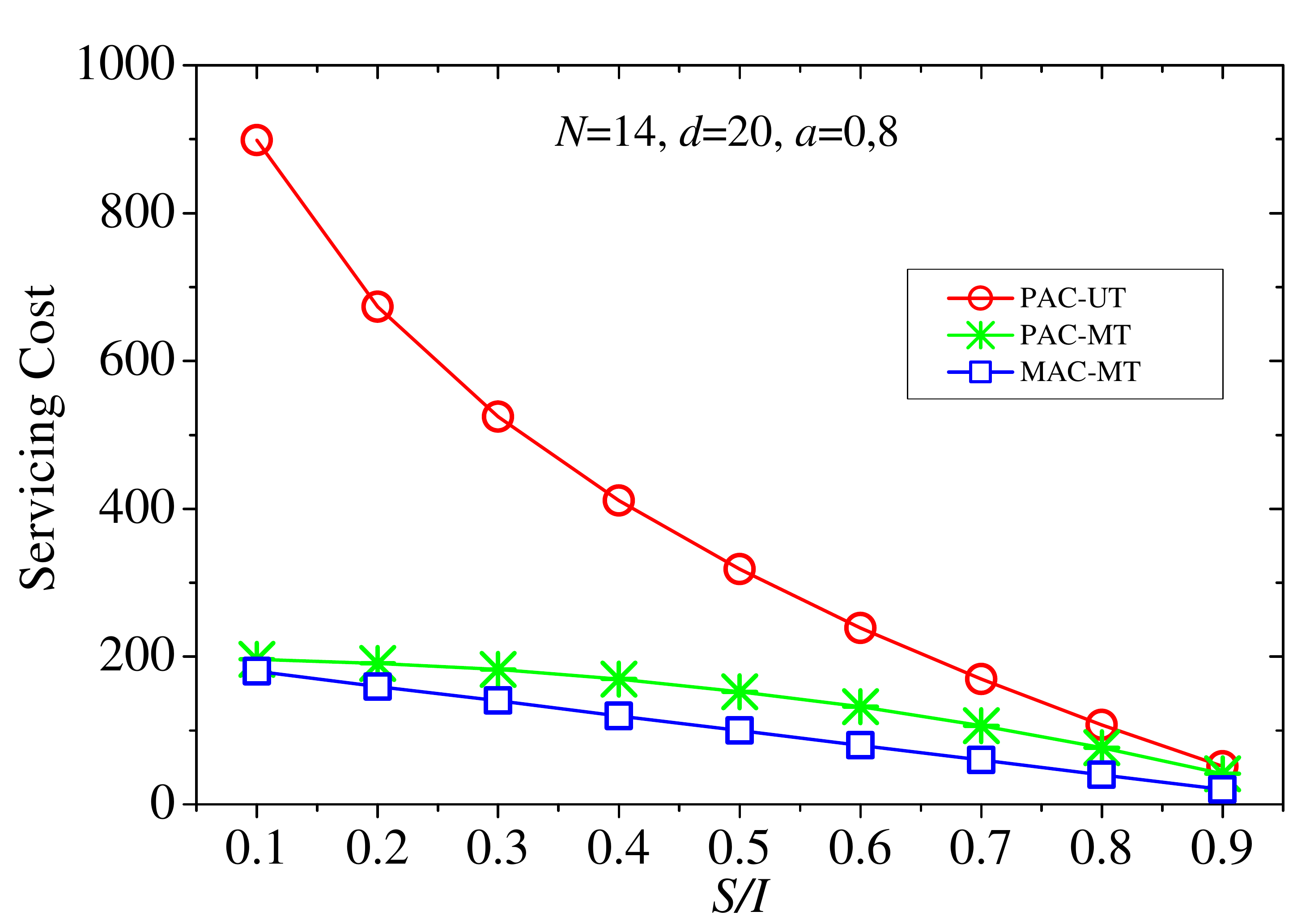}
}
\subfigure{
    \includegraphics[scale=0.17]{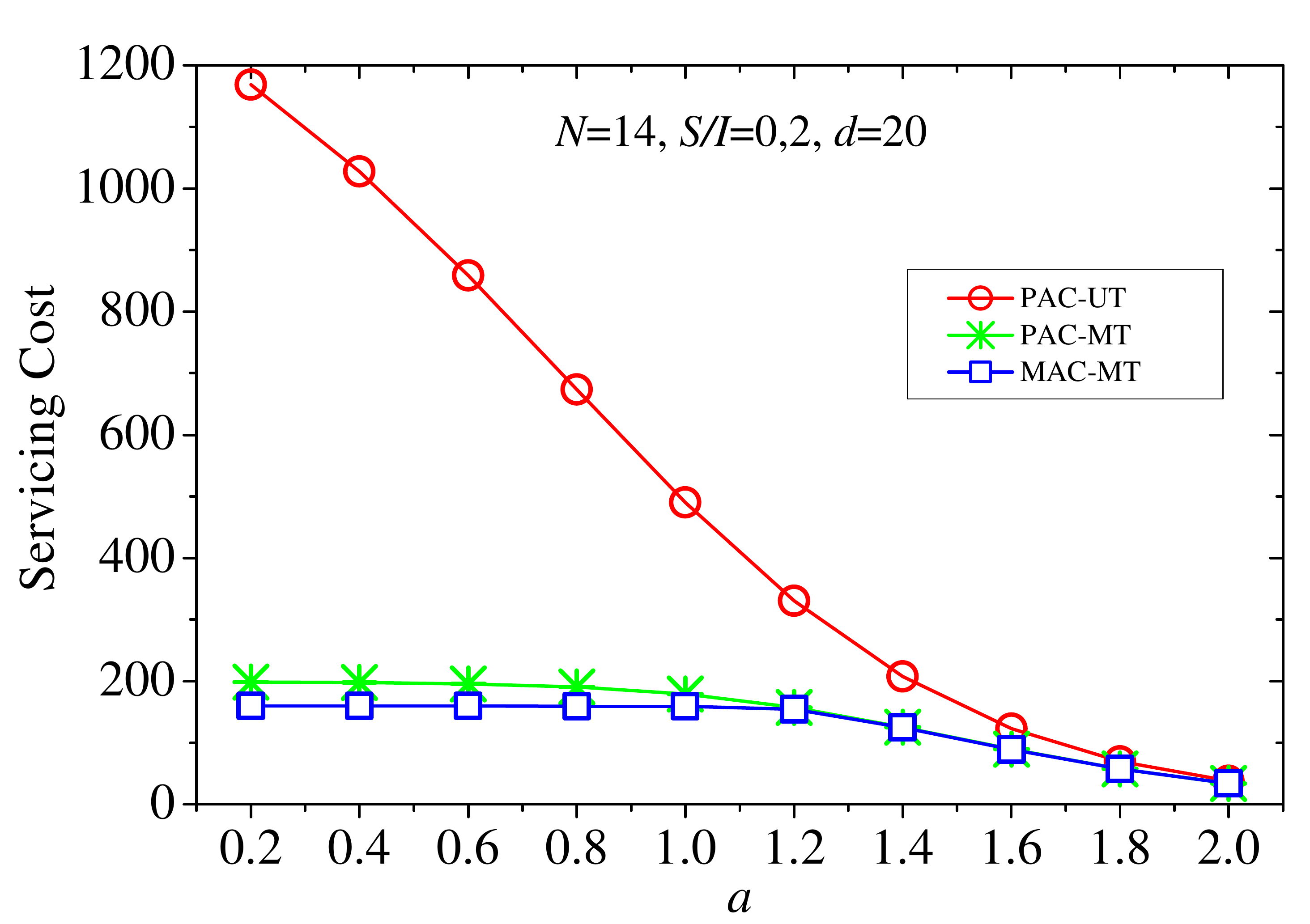}
}
\subfigure{
    \includegraphics[scale=0.17]{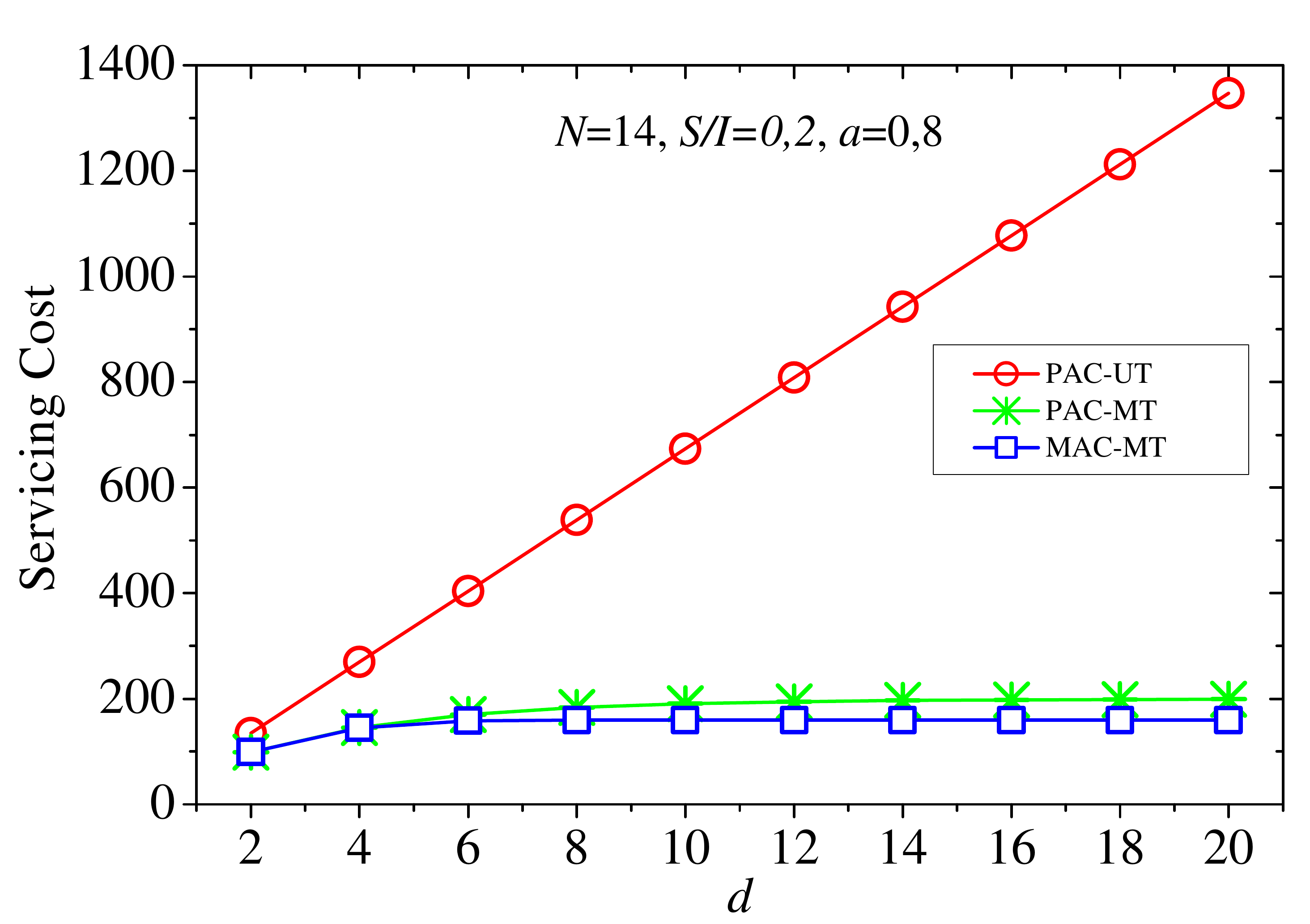}
}\\
\vspace{-2mm}
\subfigure{
    \includegraphics[scale=0.17]{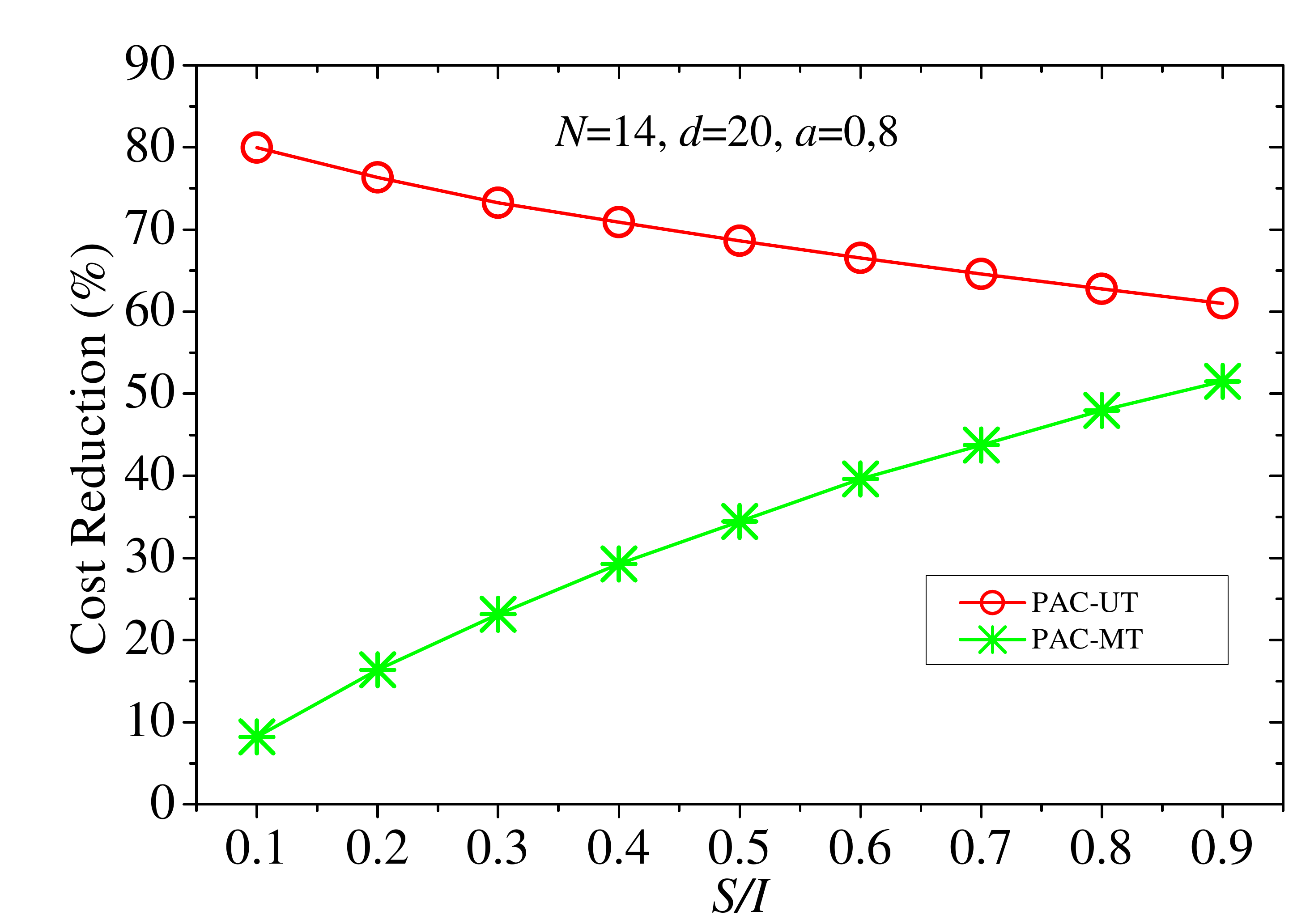}
}
\subfigure{
    \includegraphics[scale=0.17]{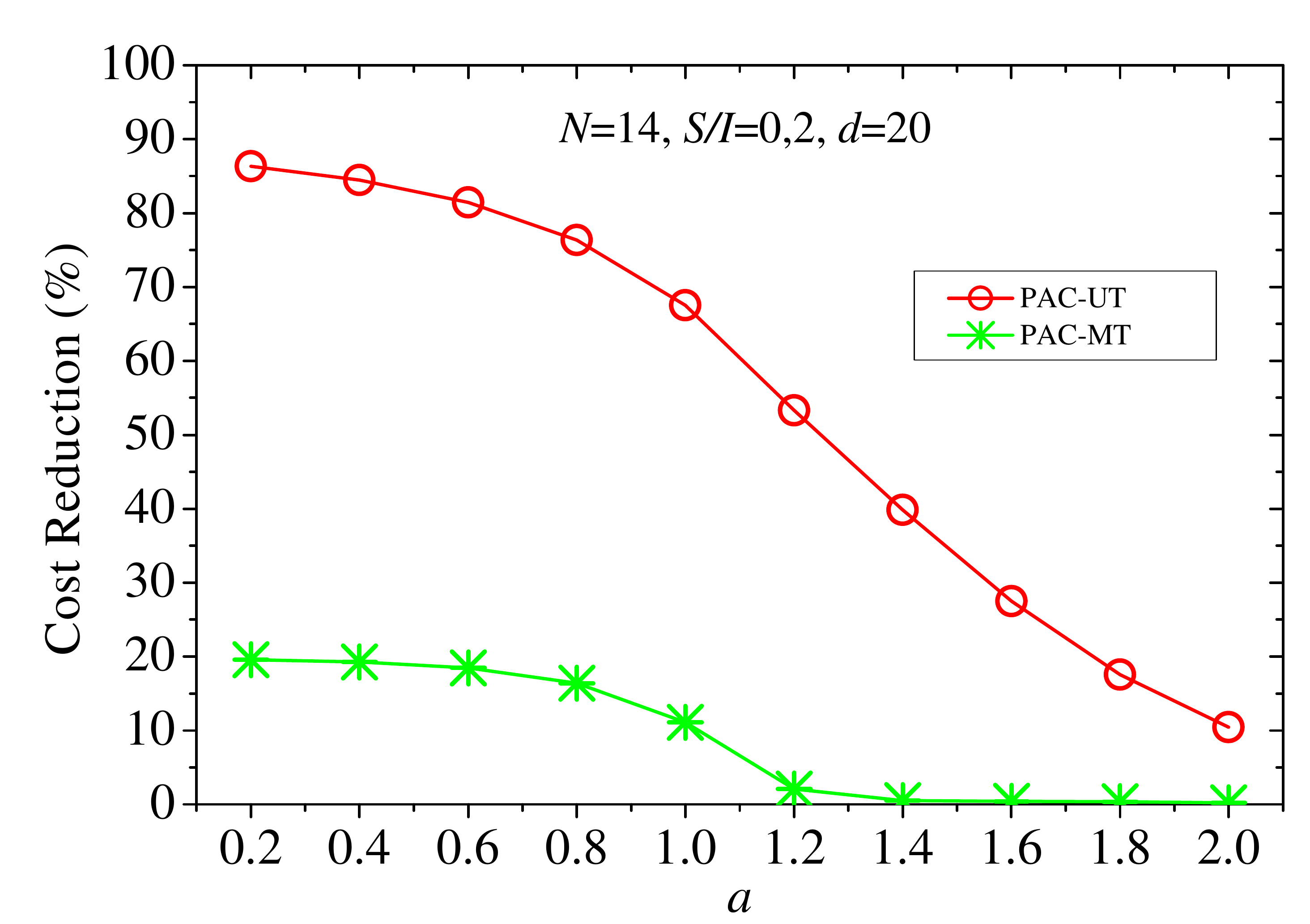}
}
\subfigure{
    \includegraphics[scale=0.17]{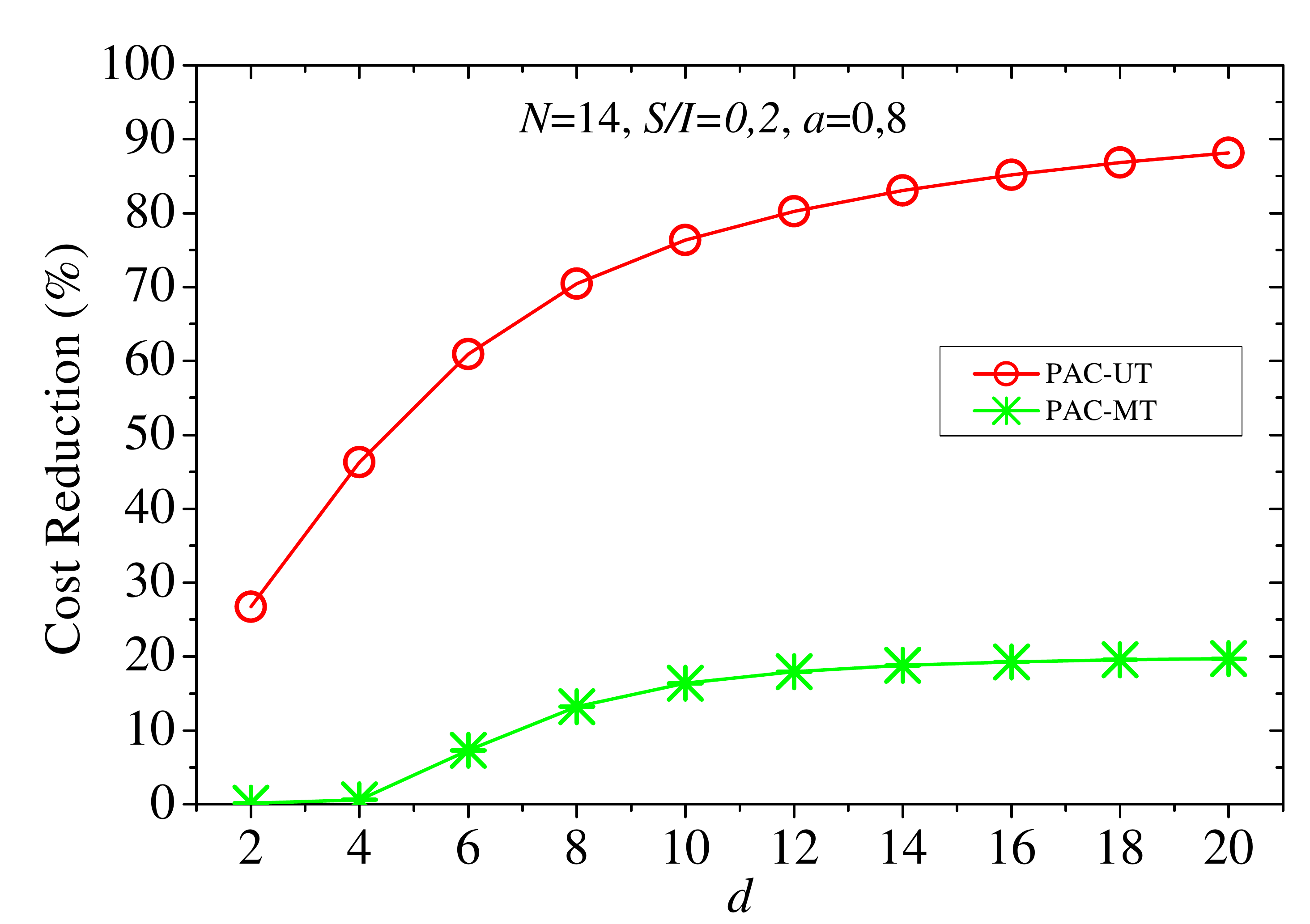}
}\\
\vspace{-1.5mm}
\text{ }\text{ } \text{ }\text{ }\text{ }\text{ }\text{ }
(a) Cache Size\text{ }\text{ }\text{ }\text{ }\text{ }\text{ }
\text{ }\text{ }\text{ }\text{ } \text{ }\text{ }
\text{ }\text{ }\text{ }\text{ }\text{ }\text{ }\text{ }\text{ }\text{ }\text{ }\text{ }
(b) Zipf-Parameter\text{ }\text{ }\text{ }
\text{ }\text{ }
\text{ }\text{ }\text{ }\text{ }\text{ }\text{ }\text{ }\text{ }\text{ }\text{ }\text{ }
\text{ }\text{ }\text{ }\text{ }\text{ }
(c) Time Deadline\text{ }\text{ }\text{ }\text{ }\text{ }\text{ }\text{ }
\begin{flushleft}
{\small
\vspace{-1.5mm}
Fig. 4. Performance comparison of PAC-UT, PAC-MT, MAC-MT for various values of (a) the cache size of each SCBS, (b) the zipf parameter of the popularity distribution of the files and (c) the time deadline.}
\end{flushleft}
\vspace{-4mm}
\end{figure*}

\section{Performance evaluation} \label{section:5}
In this section, we present the conducted experiments to evaluate the performance of the proposed algorithm.

We consider a single cell in which $N=14$ SCBSs are uniformly deployed around the MBS. We set $c_B=c_W=1$ and $c_n=0$, $\forall n \in \mathcal{N}$.
This indicates that the servicing cost comes to the operation of the MBS. Besides, recent measurement-based studies indicated that a small number of content files often accounts for a large portion of traffic \cite{DataSpotting}. We consider such a scenario and assume a limited number of $I=100$ unit-sized files. Unless otherwise specified, $S_n=S=20$ units $\forall n \in \mathcal{N}$ (i.e., can store up to $20$ files each), $d=10$ seconds, and the popularity distribution of the files follows the zipf law, with shape parameter \emph{a}$=0.8$ \cite{zipf}. The number of file requests generated within the coverage area of each SCBS follows the Poisson distribution with rate parameter that is uniformly and independently picked between the values 1 and 10 requests per second.
Finally, we set $\lambda_{0i}=0$, $\forall i \in \mathcal{I}$.

We compare the performance of three schemes:
\begin{enumerate}
\item \emph{Popularity Aware Caching $\&$ Unicast Transmissions  (PAC-UT)}: The standard mode of operation currently in use in most caching systems. Each SCBS stores in it's cache the most popular files independently from the others. Each request is served by a separate unicast transmission.
\item \emph{Popularity Aware Caching $\&$ Multicast Transmissions  (PAC-MT)}: Each SCBS stores in it's cache the most popular files independently from the others. Requests for the same file within the same time period are served by a single multicast transmission.
\item \emph{Multicast Aware Caching $\&$ Multicast Transmissions  (MAC-MT)}: We apply algorithm 1 to decide the cache placement.
    Requests for the same file within the same time period are served by a single multicast transmission.
\end{enumerate}

Below, we compare the performance of the above schemes as a function of the cache sizes, the file request pattern and the duration of the time deadline of the users.

\textbf{Impact of the cache sizes:}
Figures in 4(a) compare the performance of the discussed schemes when the cache size per SCBS is varied from $10 \%$ to $90\%$ of the entire file set size. As expected, increasing the cache sizes reduces the servicing cost of the operator as more requests are satisfied locally (without the participation of the MBS). PAC-UT results largest servicing cost compared to the other two schemes, since the latter two schemes serve many aggregated requests via a single multicast instead of many unicast transmissions. The proposed scheme (MAC-MT) consistently outperforms the others, obtaining cost reduction up to $52\%$ and $80\%$ when compared to the PAC-MT and the PAC-UT scheme respectively.

\textbf{Impact of the file request pattern:}
Figures in 4(b) illustrate the impact of the steepness of the file popularity distribution on the performance of the above schemes. We observe that as the zipf-parameter \emph{a} increases, the servicing cost decreases for all the schemes, reflecting the well known fact that caching effectiveness improves as the popularity distribution gets steeper. MAC-MT outperforms the other schemes, especially for low values of \emph{a}. The cost reduction when compared to the PAC-UT scheme is almost $88\%$ for the case that \emph{a}=0.2. However, when the popularity distribution becomes steep enough, the performance of the MAC-MT scheme is very close to the other schemes. This is because, a small number of (popular) files receive a big fraction of the users request demand, and thus the placement of these files to the caches satisfies most of the demand locally. Interestingly, in the area of \emph{a}$\in[0.6,1]$, our scheme achieves significant gains compared to the others. This is of major importance considering that the traffic generation in reality follows a zipf distribution with a parameter \emph{a} around $0.8$ \cite{zipf2}, \cite{zipf3}.

\textbf{Impact of the time deadline $d$:}
Finally, Figures 4(c) shows how the performance of the discussed schemes depends on the time parameter $d$. This is the maximum time duration that a request must be satisfied in order to be acceptable by the users (and/or the service). Particularly, as the time deadline $d$ (and hence the duration of the time period of service) becomes larger, more requests are aggregated for the same file within $d$, and thus more requests are served via multicast transmissions. Therefore, the performance gap between each one of the schemes that enable mulitcast transmissions (PAC-MT and MAC-MT) and the PAC-UT becomes larger. Besides, increasing the time deadline $d$ increases the gap between PAC-MT and MAC-MT. This is because, more multicast transmissions happen and MAC-MT is the only scheme out of the three that is designed with concerns on them.
\vspace{-1.0mm}
\section{Conclusion} \label{section:6}
\vspace{-1.0mm}
In this paper, we considered storage capable small cell base stations and proposed a novel caching scheme to minimize the cost incurred for serving the file requests of mobile users. This is a topic of major importance nowadays, as the mobile data demand growth challenges the cellular operators. In contrast to the traditional caching schemes that simply bring popular content close to the users, our caching strategy is carefully designed so as to additionally exploit the multicast opportunities. Interestingly, we find that a simple ascending greedy algorithm achieves cost reduction up to $88\%$ when compared to the existing schemes that perform only unicast transmissions. Even when multicast transmissions are employed by other schemes, our caching policy outperforms them, achieving cost reduction up to $52 \%$.

\end{document}